\documentstyle[12pt]{article}
\input epsf
\title{\bf Anyonic behavior of quantum group gases}
\author {\\ \\ \\ Marcelo R. Ubriaco\thanks{E-mail:ubriaco@ltp.upr.clu.edu}\\
{\em Laboratory of Theoretical Physics}\\
{\em Department of Physics}\\
{\em  University of Puerto Rico}\\
{\em P. O. Box 23343, R\'{\i}o Piedras}\\
{\em PR 00931-3343, USA}}
\date{}
\begin{document}
\vspace{0.3in}
\maketitle
\vspace{0.15in}
\begin{abstract}
We  first introduce and discuss the formalism of $SU_q(N)$-bosons and
fermions and  consider the simplest Hamiltonian
involving these operators.  We then calculate the
grand partition function for these models and study the high temperature (low density) case 
of the
corresponding gases for $N=2$. We show that quantum group gases
exhibit anyonic behavior in $D=2$ and $D=3$ spatial dimensions.
In particular, for a $SU_q(2)$ boson gas at $D=2$  the parameter $q$
interpolates within a wider range of attractive and repulsive
systems than the anyon statistical parameter. 
\end{abstract}

\newpage
\baselineskip20pt
\section{Introduction}
In the last few years, the search for new applications of quantum groups and quantum algebras
\cite{Jimbo,Ch}, other than the theory of integrable models and the
quantum inverse scattering method, has attracted 
 the attention of mathematicians and physicists alike.
The published literature on  formulations based 
on quantum group theory includes studies in
non-commutative geometry \cite{Wo,WZ}, quantum mechanics \cite{Wat},
field theory \cite{AV}, molecular and nuclear physics \cite{I}.
Many of these approaches are attempts to develop
more general formulations of quantum mechanics and field
theory, and to look for small deviations from
the standard value $q=1$ in nuclear and molecular physics.
In this article we study the high temperature
(low density) behavior of two quantum group gases.
In Section \ref{qgbf} we  discuss
the covariant $SU_q(N)$ fermion and
boson algebras, and specialize to the case $N=2$.
In subsections \ref{fm} and \ref{fm} we introduce
the $SU_q(2)$ fermion and boson models respectively,
and in each case we give a representation of these operators
in terms of the corresponding standard fermion or boson
oscillators. Section \ref{qgg} contains the main results of this
work. 
We obtain the equation of  state  as a virial
expansion and
discuss their anyonic behavior for both gases at $D=2$ and $D=3$. In $D=2$ we compare the
parameter $q$ with the anyon statistical parameter $\alpha$.
\section{Quantum group bosons and fermions}\label{qgbf}
In this section we briefly discuss the quantum group field algebras
introduced in Reference \cite{U1}.  These algebras can be
seen as  generalizations of the standard bosonic and
fermionic algebras.  As it is well known, bosonic and fermionic operators
satisfy the algebraic relations
\begin{eqnarray}
\phi_i\phi_j^{\dagger}-\phi_j^{\dagger}\phi_i&=&\delta_{ij}\nonumber\\
\psi_i\psi_j^{\dagger}+\psi_j^{\dagger}\psi_i&=&\delta_{ij}, 
\end{eqnarray}
which, for $i,j=1,...N$, are covariant under $SU(N)$ transformations.  
The quantum group analogues of these equations are given by the
 following relations            
\begin{equation}
\Omega_j\overline{\Omega}_i=\delta_{ij}\pm q^{\pm1}R_{kijl}
\overline{\Omega}_l\Omega_k \label{c1}
\end{equation}
\begin{equation}
\Omega_l\Omega_k=\pm q^{\mp 1}R_{jikl}\Omega_j\Omega_i,\label{c2}
\end{equation}
where $\Omega=\Phi,\Psi$ and the upper (lower) sign applies to quantum group bosons $\Phi_i$
(quantum group fermions $\Psi_i$) operators.  The $N^2\times N^2$ matrix 
 $R_{jikl}$ is explicitly written \cite{WZ}
\begin{equation}
R_{jikl}=\delta_{jk}\delta_{il}(1+(q-1)\delta_{ij})
+(q-q^{-1})\delta_{ik}\delta_{jl}\theta(j-i),
\end{equation}
where $\theta(j-i)=1$ for $j>i$ and zero otherwise. Denoting 
the new fields as $\Omega_i'=\sum_{i=1}
^{N}T_{ij}\Omega_j$, the
$SU_{q}(N)$ transformation matrix $T$ 
and the $R$-matrix  
satisfy the well known algebraic relations \cite{Ta}
\begin{equation}
RT_1T_2=T_2T_1R,\label{T}
\end{equation}
and
\begin{equation}
R_{12}R_{13}R_{23}=R_{23}R_{13}R_{12},
\end{equation}
with the standard embedding $T_1=T\otimes 1$, $T_2=1\otimes T$
$\in V\otimes V$ and $(R_{23})_{ijk,i'j'k'}=
\delta_{ii'} R_{jk,j'k'} \in V\otimes V\otimes V$.

In particular, for $N=2$, Equations (\ref{c1}) and (\ref{c2}) 
are simply written
\begin{description}
\item[a)] $SU_q(2)-fermions$
\begin{eqnarray}
\{\Psi_{2},\overline{\Psi}_{2}\}&=&1\\ \label{f0}
\{\Psi_{1},\overline{\Psi}_{1}\}&=&1 - (1-q^{-2})\overline{\Psi}_{2}\Psi_{2}\label{f1}\\ 
\Psi_{1}\Psi_{2}&=&-q \Psi_{2}\Psi_{1}\\ 
\overline{\Psi}_{1}\Psi_{2}&=&-q \Psi_{2}\overline{\Psi}_{1}\\
\{\Psi_{1},\Psi_{1}\}&=&0=\{\Psi_{2},\Psi_{2}\} \label{0},\label{01}
\end{eqnarray}
\item[b)] $SU_q(2)-bosons$
\begin{eqnarray}
\Phi_2\overline{\Phi}_2-q^2\overline{\Phi}_2\Phi_2&=&1 \\
\Phi_1\overline{\Phi}_1-q^2\overline{\Phi}_1\Phi_1&=&1+(q^2-1)\overline{\Phi}_2\Phi_2
\\
\Phi_2\Phi_1&=&q\Phi_1\Phi_2\\
\Phi_2\overline{\Phi}_1&=&q\overline{\Phi}_1\Phi_2
\end{eqnarray}
which for $q=1$ become the fermion and boson algebras respectively.
According to Equation (\ref{T}) the matrix $T=\left(\begin{array}{cc} a & b \\ c & d\end{array}\right)$
elements generate the algebra
\begin{eqnarray}
ab=q^{-1}ba  & , & ac=q^{-1}ca \nonumber \\
bc=cb & , & dc=qcd  \nonumber \\
db=qbd & , &  da-ad=(q-q^{-1})bc  \nonumber \\
& & det_{q}T\equiv ad-q^{-1}bc=1 ,
\end{eqnarray} 
with the unitary conditions \cite{VWZ} $\overline{a}=d, \overline{b}=q^{-1}c$
and $q\in {\bf R}$. Hereafter, we take $0\leq q<\infty$.
\end{description}

\subsection{$SU_q(2)$ fermion model}\label{fm}
The simplest Hamiltonian one can write in terms of
the operators $\Psi_i$ is simply 
\begin{equation}
{\cal H}_F=\sum_{\kappa}^{}\varepsilon_{\kappa}({\cal M}_{1,\kappa}+
{\cal M}_{2,\kappa}),\label{h}
\end{equation}
where ${\cal M}_{i\kappa}=\overline{\Psi}_{i,\kappa}\Psi_{i,\kappa}$
and $\{\overline{\Psi}_{\kappa,i},\Psi_{\kappa',j}\}=0$ for $\kappa\neq\kappa'$.
From Equation (\ref{01}) we see that the occupation numbers are restricted to $m=0,1$ and 
therefore $SU_q(N)$ fermions satisfy the Pauli exclusion principle. 
For a given $\kappa$ a normalized state is simply written
\begin{equation}
\overline{\Psi}_2^n\overline{\Psi}_1^m|0>,\;\;\; n,m=0,1 \label{s}
\end{equation}
 and the ${\cal M}_i$  operator satisfies
\begin{equation}
[{\cal M}_2,\Psi_1]=0 ,
\end{equation}
and
\begin{equation}
{\cal M}_1\Psi_2-q^2\Psi_2{\cal M}_1=0.
\end{equation}
The grand partition function is given by
\begin{equation}
{\cal Z}_F= Tr\; e^{-\sum_\kappa \varepsilon_\kappa({\cal M}_{1,\kappa}+{\cal M}_{2,\kappa})}
e^{\beta\mu(M_{1,\kappa}+M_{2,\kappa})},
\end{equation}
where $M_{i,\kappa}=\psi_{i,\kappa}^{\dagger}\psi_{i,\kappa}$ 
are the standard fermion number operators, and the trace is taken with
respect to the states in Equation (\ref{s}).  Since the pair $\Psi_2,\overline{\Psi}_2$
satisfies standard anticommutation relations we can identify it without 
any loss of generality with a fermion pair $\psi_2,\psi_2^\dagger$. In addition,
from Equations (\ref{f1}) and (\ref{01}) we see that the operator $\Psi_1(\overline{\Psi}_1)$
is a function of the operator $\psi_1(\psi_1^{\dagger})$ times a function of
 ${\cal M}_2$.  Therefore
the grand partition function ${\cal Z}_F$ becomes
\begin{eqnarray}
{\cal Z}_F&=&\prod_\kappa\sum_{n=0}^1\sum_{m=0}^1e^{-\beta\varepsilon_\kappa(n+m
-(1-q^{-2})mn}e^{\beta\mu(n+m)}\\
&=&\prod_\kappa \left(1+2e^{-\beta(\varepsilon_\kappa-\mu)}+e^{-\beta\left
(\varepsilon_\kappa(q^{-2} +1)-2\mu\right)}\right),\label{ZF}
\end{eqnarray}
which for $q=1$ becomes the square of a single fermion type grand partition
function.  From Equation (\ref{ZF}) we see  that the original Hamiltonian becomes 
 the interacting 
Hamiltonian
\begin{equation}
H_F=\sum_\kappa \varepsilon_\kappa\left(M_{1,\kappa}+M_{2,\kappa}+(q^{-2}-1)
M_{1,\kappa}M_{2,\kappa}\right).\label{HF}
\end{equation}
Therefore the parameter $q\neq 1$ mixes the two
degrees of freedom in a nontrivial way through a quartic term in the Hamiltonian.
The thermodynamics of this system will be discussed in section (\ref{thf})

A simple check shows that Equations (\ref{f0})-(\ref{01}) and (\ref{HF})
are consistent with the following representation of $\Psi_i$ operators
in terms of fermions operators $\psi_j$ 
\begin{eqnarray}
\Psi_2&=&\psi_2\\
\overline{\Psi}_2&=&\psi_2^{\dagger}\\
\Psi_1&=&\psi_1\left(1+(q^{-1}-1)M_2\right)\\
\overline{\Psi}_1&=&\psi_1^{\dagger}\left(1+(q^{-1}-1)M_2\right),
\end{eqnarray}
and according to Equations (\ref{c1}) and (\ref{c2}) this result easily
generalizes for arbitrary $N$ to
\begin{equation}
\Psi_m=\psi_m\prod_{l=m+1}^{N}\left(1+(q^{-1}-1)M_l\right),
\end{equation}
and similarly for the adjoint equation.

It is interesting to remark the distinction between $SU_q(2)$-fermions
with the so called $q$-fermions $b_i$ and $b_i^\dagger$. The $q$-fermionic
 algebra was introduced in \cite{NG}
\begin{eqnarray}
bb^\dagger+qb^\dagger b&=&q^{N_q}\\
b^\dagger b&=&[N_q]\\
bb^\dagger&=&[1-N_q]\\
b^2=&0&=b^{\dagger 2},
\end{eqnarray}
where the bracket $[x]=\frac{q^x-q^{-x}}{q-q^-1}$ and the number operator
$N_q|n\rangle=n|n\rangle$ with $n=0,1$.  Since the $q$-number $[x]=x$ for
$x=0,1$, it is obvious that the grand partition function for $q$-fermions is
no different than the Fermi grand partition function, and therefore the
$q$-fermions do not lead to new results as far as thermodynamics is concerned.
\subsection{$SU_q(2)$ boson model}\label{bm}
In terms of $SU_q(2)$-bosons we introduce the following Hamiltonian
\begin{equation}
{\cal H}_B=\sum_\kappa \varepsilon_\kappa({\cal N}_{1,\kappa}+{\cal N}_{2,\kappa}),
\end{equation}
where $[\overline{\Phi}_{i,\kappa},\Phi_{\kappa',j}]=0$ for
$\kappa\neq\kappa'$.
The operator ${\cal N}_{i,\kappa}=\overline{\Phi}_{i,\kappa}\Phi_{i,\kappa}$
satisfy the relations
\begin{equation}
[{\cal N}_{2,\kappa},\Phi_1]=0,
\end{equation}
and
\begin{equation}
{\cal N}_{1,\kappa}\Phi_2-q^{-2}\Phi_2{\cal N}_{1,\kappa}=0.
\end{equation}
 The states are built by the action of the $\Phi$
operators on the vacuum state.  For example, for a given $\kappa$
a normalized state with $n_1$ particles of species $1$ and $n_2$ particles
of species $2$ is defined by
\begin{equation}
\frac{1}{\sqrt{\{n_1\}!\{n_2\}!}}\overline{\Phi}_2^{n_2}
\overline{\Phi}_1^{n_1}|0\rangle,\label{b}
\end{equation}
where  the $q$-numbers $\{n\}=\frac{1-q^{2n}}{1-q^2}$ and the $q$-factorials $\{n\}!$
are defined $\{n\}!=\{n\}\{n-1\}\{n-2\}...1$. 
The grand partition function ${\cal Z}_B$ is written
\begin{equation}
{\cal Z}_B= Tr\; e^{-\beta\varepsilon_\kappa(\overline{\Phi}_{1,\kappa}\Phi_{1,\kappa}
+\overline{\Phi}_{2,\kappa}\Phi_{2,\kappa})}e^{-\beta\mu(N_{1,\kappa}+N_{2,\kappa})}	,
\end{equation}
where $N_{i,\kappa}$ are the ordinary boson number operators
$N_{i,\kappa}=\phi_{i,\kappa}^\dagger \phi_{i,\kappa}$ and the trace is taken
with respect to the states in Equation (\ref{b}).  
For a given $\kappa$ the $SU_q(2)$ bosons are written 
in terms of boson operators $\phi_{i,\kappa}$
and $\phi_{i,\kappa}^{\dagger}$ with usual commutations relations
$[\phi_i,\phi_j^{\dagger}]=\delta_{ij}$ as follows
\begin{eqnarray}
\Phi_2&=&(\phi_2^\dagger)^{-1} \{N_2\}\label{b1}\\
\overline{\Phi}_2&=&\phi_2^\dagger \\
\Phi_1&=&(\phi_1^\dagger)^{-1} \{N_1\}q^{N_2}\\
\overline{\Phi}_1&=&\phi_1^\dagger q^{N_2}\label{b4}
\end{eqnarray}

The grand partition function ${\cal Z}_B$ then
becomes 
\begin{equation}
{\cal Z}_B=\prod_\kappa\sum_{n=0}^{\infty}\sum_{m=0}^{\infty}
e^{-\beta\varepsilon_\kappa\{n+m\}}e^{\beta\mu(n+m)},\label{Zb}
\end{equation}
with the corresponding interacting Hamiltonian 
\begin{equation}
H_B=\sum_{\kappa}\varepsilon_{\kappa}\{\phi_{1,\kappa}^{\dagger}
\phi_{1,\kappa}+\phi_{2,\kappa}^{\dagger}
\phi_{2,\kappa}\},
\end{equation}
with the bracket $\{x\}$ as defined below Equation (\ref{b}).
Therefore, the original Hamiltonian 
becomes a Hamiltonian in terms of ordinary boson 
interactions involving powers of the number operators $N_{i,\kappa}$ and
$\log q$.  Equations (\ref{b1})-(\ref{b4}) are easily generalized
for $N>2$ to the set of equations
\begin{equation}
\overline{\Phi}_m=\phi_m^{\dagger}\prod_{l=m+1}^N q^{N_l},
\end {equation}
and
\begin{equation}
\Phi_m=(\phi_m^{\dagger})^{-1}\{N_m\}\prod_{l=m+1}^N q^{N_l},
\end{equation}
and  a $SU_q(N)$-boson state in terms of boson operators reads
\begin{equation}
\frac{1}{\sqrt{\{n_1\}!\{n_2\}!...\{n_M\}!}}\phi_{M,\kappa_M}
^{\dagger n_M}\phi_{M-1,\kappa_{M-1}}^{\dagger n_{M-1}}
...\phi_{1,\kappa_1}^{\dagger n_1}|0\rangle.\label{sb}
\end{equation}
The normalization is consistent with the fact that the dual of the
state in Equation (\ref{sb}) is obtained 
by applying the adjoint operation defined on $\Phi$.
The number operator $N_l=\phi_l^{\dagger}\phi_l$ satisfies standard
commutation relations with the operators $\Phi_m$
\begin{equation}
[N_{l,\kappa},\overline{\Phi}_{m,\kappa'}]=\overline{\Phi}_{m,\kappa'}
\delta_{\kappa,\kappa'} \delta_{l,m}
\end{equation}
and
\begin{equation}
[N_{l,\kappa},\Phi_{m,\kappa'}]=-\Phi_{m,\kappa}\delta_{\kappa,\kappa'}
\delta_{l,m},
\end{equation}
such that
\begin{equation}
N_l\overline{\Phi}_l^m|0\rangle=m\overline{\Phi}_l^m|0\rangle.
\end{equation}
The difference between the operators $\Phi$ and the so called
$q$-bosons  is obvious.  A set $(a_i,a_i^{\dagger})$ of $q$-bosons
satisfies the relation \cite{Mf,B}
\begin{equation}
a_i a_i^{\dagger}-q^{-1}a_i^{\dagger}a_i=q^N ,\;\;\;
[a_i,a_j^{\dagger}]=0=[a_i,a_j],\label{q1}
\end{equation}
where $N|n\rangle=n|n\rangle$. By taking two commuting sets
of $q$-bosons it has been shown \cite{NG} that the operators
\begin{equation}
J_+=a_2^{\dagger}a_1\;,\;\;J_-=a_1^{\dagger}a_2\;,\;\;
2J_3=N_2-N_1
\end{equation}
generate the quantum Lie algebra $su_q(2)$
\begin{equation}
[J_3,J_{\pm}]=\pm J_{\pm}\;,\;\;\;\;
[J_-,J_-]=[2J_3].
\end{equation}
In contrast to the algebraic relations
involving the operators $\Phi_i$ and $\overline{\Phi}_j$, Equation
(\ref{q1}) with $i,j=1,2$ is not covariant under 
the action of the $SU_q(2)$ quantum group matrices.  The thermodynamics
of $q$-bosons and similar operators called quons
\cite{G} has been studied by several authors \cite{TH}\cite{An}.  In the following
section we study the thermodynamics of the two $SU_q(2)$ models described in
this section .
\section{Quantum group gases}\label{qgg}
The high and low temperature behavior of the $SU_q(2)$ fermion model  
has been studied in References \cite{U2,U3}, and here we recall
some results that will be compared with the $SU_q(2)$ boson case.
\subsection{Quantum group fermion gas}\label{thf}
 The internal energy $U$ for this model is calculated
 from the grand potential $\Omega=-\frac{1}{\beta}\ln {\cal Z}_F$
according to the equation
\begin{eqnarray}
U&=&(\frac{\partial\beta\Omega}{\partial\beta}+\mu M)\nonumber\\
&=&V\int_{}^{}\frac{p^{2}}{2m}\frac{
(2+(q^{-2}+1)e^{\beta(\mu-\frac{q^{-2}p^{2}}{2m}})d^{3}p}{(2\pi\hbar)^{3}
f(\varepsilon,\mu,q)},\label{U}
\end{eqnarray} 
where the function $f(\varepsilon,\mu,q)=e^{\beta(\varepsilon-\mu)}+2
+e^{-\beta(q^{-2}\varepsilon-\mu)}$.
The low temperature regime of a $SU_q(2)$ fermion gas exhibits
the interesting feature that for every value of $q\neq 1$ the
entropy lies below the Fermi entropy. For $q>1$ and $q<1$ the
entropy
functions are given respectively by the equations \cite{U3}
\begin{equation}
S(q>1)\approx\lambda\frac{1.28\sqrt{2\mu_{0}}k^{2}T}{(q^{-2}+1)^{3/2}},\label{S+}
\end{equation}
and
\begin{equation}
S(q<1)\approx\lambda k^{2}\sqrt{\mu_{0}}T\left[1.08(q^{3}+1)
-\frac{(1-q^{3})^{2}}{2(1+q^{3})}\ln^{2}3\right]\label{S-},
\end{equation}
where $\lambda=\frac{4\pi V(2m)^{3/2}}{(2\pi\hbar)^{3}}$.
The lower bound to the
entropy values corresponds to the limit $q\rightarrow 0$.  Furthermore,
 systems described by a  Hamiltonian with $q>1$ share the same entropy
function with systems with $q<1$. Comparing Equation (\ref{S+}) with Equation (\ref{S-})
we obtain that two gases share the
entropy function if the following relation is satisfied
\begin{equation}
(1+q'^{-2})^{3/2}=\frac{3.62(1+q^3)}{2.16(1+q^3)^2-(1-q^3)^2\ln^2 3},
\end{equation}
where $q'>1$ and $q<1$. Specifically, the equality is satisfied
in the interval $0.33\leq q<0.91$.

 The high temperature behavior 
of this model is also interesting. Starting with the grand partition
function ${\cal Z}_F$
\begin{equation}
\ln {\cal Z}_F=\frac{4\pi V}{h^3}\int_0^{\infty}
p^2\ln\left(1+2e^{-\beta(\varepsilon-\mu)}+e^{-\beta\left(\varepsilon
(q^{-2}+1)-2\mu\right)}\right) dp
\end{equation}
it was shown in Reference
\cite{U3} that in $D=3$ spatial dimensions the virial expansion leads to the equation of state
\begin{equation}
pV=kT\langle M\rangle\left(1+\frac{\alpha(q)}
{2}\left(\frac{h^2}{2\pi mkT}\right)^{3/2}\frac{\langle M\rangle}{V}+...\right),\label{pv}
\end{equation}
where the coefficient $\alpha(q)=\frac{1}{2^{3/2}}-\frac{1}{2(q^{-2}+1)^{3/2}}$.
From equation (\ref{pv}) we see that the sign of the
second virial coefficient depends on the value of $q$, showing 
that the parameter $q$ interpolates between attractive and repulsive
behavior.  The function $\frac{\alpha}{2}$ takes values in the
interval $ 2^{-5/2}\geq\frac{\alpha}{2}\geq -2^{-5/2}(\sqrt{2}-1)$ as $q$ varies
from zero to $\infty$ and vanishes at $q=1.96$. 
Figure 1 shows a graph of the function $B(q,T)=\frac{\alpha(q)}{2}\beta^{3/2}$
for large values  of the temperature and $q=10,1.96,1,0.3$.

It is important to remark that
the free boson limit $B_b(T)=-2^{-7/2}\beta^{3/2}<B(\infty,T)=-2^{-5/2}(\sqrt{2}-1)\beta^{3/2}$,
and therefore free bosons are not described in this model. A natural
question to address is whether a similar interpolation occurs
at $D=2$. The same procedure leads to the
equation of state
\begin{equation}
pA=kT\langle M\rangle\left(1+\frac{1}{1+q^2}\frac{h^2}{8\pi mkT}
\frac{\langle M\rangle}{A}+...\right),
\end{equation}
wherein  the second virial coefficient is positive for all values
of $q$, showing  that this model
,at $D=2$, describes only repulsive systems. 
\subsection{Quantum group boson gas}
The grand partition function ${\cal Z}_B$ in Equation (\ref{Zb})
can be simply rewritten as
\begin{equation}
{\cal Z}_B=\prod_\kappa\sum_{m=0}^{\infty} (m+1)e^{-\beta\varepsilon_{\kappa}\{m\}}z^m,
\end{equation}
where $z=e^{\beta\mu}$ is the fugacity.  In $D=3$ the first few terms in powers of $z$ read
\begin{eqnarray}
\ln{\cal Z}_B&=&\frac{4\pi V}{h^3}\int_{0}^{\infty}dp p^2( 2e^{-\beta\varepsilon_{\kappa}}z
+(6 e^{-\beta\varepsilon_{\kappa}\{2\}}-4e^{-\beta\varepsilon_{\kappa}2})\frac{z^2}{2}\nonumber\\
&+&(24 e^{-\beta\varepsilon_{\kappa}\{3\}}-36 e^{-\beta\varepsilon_{\kappa}\{2\}}
e^{-\beta\varepsilon_{\kappa}}
+16 e^{-\beta\varepsilon_{\kappa}3})\frac{z^3}{3!}+...),
\end{eqnarray}
such that performing the elementary integrations gives
\begin{equation}
\ln {\cal Z}_B=\frac{4\pi V}{h^3}\left(\frac{\sqrt{\pi}}{2}(\frac{2m}{\beta})^{3/2} z+
\sqrt{\pi}(\frac{2m}{\beta})^{3/2}\delta(q) z^2+...\right),
\end{equation}
where $\delta(q)=\frac{1}{4}\left(\frac{3}{(1+q^2)^{3/2}}-\frac{1}{\sqrt{2}}\right)$. 

Calculating the
average
 number of particles $\langle N\rangle=\frac{1}{\beta}\left(\frac{\partial\ln{\cal Z}_B}{\partial\mu}\right)
_{T,V}$ and reverting the equation we find for the fugacity 
\begin{equation}
z\approx \frac{1}{2}\left(\frac{h^2}{2m\pi kT}\right)^{3/2}
\frac{\langle N\rangle}{V}-\delta(q) \left(\frac{h^2}{2m\pi kT}\right)^{3}\left(\frac{\langle N
\rangle}{V}\right)^2,
\end{equation}
leading to the following equation of state
\begin{equation}
pV=kT\langle N\rangle\left(1-\delta(q)\left(\frac{h^2}{2m\pi kT}\right)^{3/2}\frac{\langle N\rangle}{V}
+...\right).
\end{equation}

As expected, at $q=1$ the coefficient $\delta(1)=2^{-7/2}$, which is the numerical
factor in the second virial coefficient for a free boson gas with two species.
The free fermion $\delta(q)=-2^{7/2}$ and ideal gas $\delta(q)=0$
 cases   are reached at  $q\approx 1.78$
and $q\approx 1.27$ respectively.

A very similar calculation for $D=2$ gives the equation of state
\begin{equation}
pA=kT\langle N\rangle\left(1-\eta(q)\frac{h^2}{2\pi mkT}\frac{\langle N\rangle}{A}+...\right),\label{pA}
\end{equation}
with $\eta(q)=\frac{(2-q^2)}{4(1+q^2)}$. At $D=2$ this model behaves as a
 fermion gas for $q=\sqrt{5}$.  Figure 2 shows a graph of the coefficient $\eta(q)$
 as a function of the parameter
$q$ for $D=2$.

Since the $SU_q(2)$
boson gas at $D=2$ also interpolates completely between bosons and fermions,
 we can find a relation
between the parameter $q$ and the  statistical parameter $\alpha$ for an
anyon gas \cite{W,A} of two species. This relation is given by
\begin{equation}
\alpha=1-\sqrt{\frac{5-q^2}{2(1+q^2)}},
\end{equation}
where $0\leq\alpha\leq 1$, with the boson and fermion limits $\alpha=0$ $(q=1)$
and $\alpha=1$ $(q=\sqrt{5})$ respectively.  The second virial 
coefficient in Equation (\ref{pA})
takes values in the interval $[-\frac{\lambda_T^2}{2},\frac{\lambda_T^2}{4}]$, with 
$\lambda_T=\sqrt{\frac{h^2}{2\pi mkT}}$, and therefore the parameter $q$
interpolates within a larger range of systems than the $\alpha$ parameter does.
\section{Conclusions}
In this article we have 
studied the high temperature behavior of quantum group gases.
Our approach is mainly based on promoting the $SU(N)$
covariant fermion and boson algebras to the
corresponding algebraic
relations covariant under $SU_q(N)$ transformations.
For purposes of simplicity we have considered the $N=2$ case.
Starting with the simplest Hamiltonian we have calculated
the partition function and obtained
the equation of state for the two $SU_q(2)$ gases.
Certainly, for $q=1$ our results become those for two species of free fermion or
boson gases. For $q\neq 1$ this degeneracy is broken and 
the corresponding Hamiltonian written in terms of
standard operators acquires an interaction term.
Our results indicate that 
 the $q$ parameter interpolates between repulsive and attractive behavior.
In particular, for a $SU_q(2)$-fermion gas and D=3 the sign of the second virial
coefficient depends on the value of $q$. The ideal gas case corresponds
to $q=1.96$ and the system becomes repulsive for $q<1.96$. For $q>1.96$
the system becomes attractive, but as
$q\rightarrow\infty$ the free boson limit is not reached, and therefore
this model does not interpolate completely between the free fermion and
free boson cases. For $D=2$ the second
virial coefficient of this gas is positive for every value of $q$ and
vanishes in the $q\rightarrow\infty$ limit.  
For $SU_q(2)$-bosons the results are more interesting. For $D=2$ and $D=3$
the parameter $q$ interpolates completely between a wide  range of 
attractive and repulsive systems including the free fermion and boson
cases.  For $D=2$ we have found a relation between $q$ and the
statistical parameter $\alpha$ for an anyon gas. Therefore, 
the simple models  studied here, and in particular the
$SU_q(2)$-boson model, offer an alternative approach
in describing systems obeying fractional statistics in two and
three spatial dimensions.

\end{document}